\RequirePackage{fixltx2e}
\documentclass[twocolumn,secnumarabic,superscriptaddress,amssymb, nobibnotes, aps, prb]{revtex4-1}
\usepackage[figuresright]{rotating}
\usepackage[T1]{fontenc}
\usepackage{layouts}
\usepackage{amsfonts}
\usepackage{dcolumn}
\usepackage{amssymb}
\usepackage{amsmath}
\usepackage{graphicx}
\usepackage{supertabular}
\usepackage{bm}
\usepackage{dsfont}
\usepackage{braket}
\usepackage{hyperref}
\usepackage{color,soul}
\usepackage[english]{babel}

\let\oldAA\AA
\renewcommand{\AA}{\text{\normalfont\oldAA}}

\begin{document}

\title{Spin-polarized electronic surface states of Re(0001): an ab-initio investigation}%

\author{Andrea Urru}
\affiliation{International School for Advanced Studies (SISSA), \\
Via Bonomea 265, 34136 Trieste (Italy).}
\author{Andrea Dal Corso}
\affiliation{International School for Advanced Studies (SISSA), \\
Via Bonomea 265, 34136 Trieste (Italy).}
\affiliation{IOM-CNR Trieste (Italy).}

\begin{abstract}
We study the electronic structure of the Re(0001) surface by means of ab-initio
techniques based on the Fully Relativistic (FR) Density Functional Theory (DFT)
and the Projector Augmented-Wave (PAW) method.
We identify the main surface states and resonances and study in detail their
energy dispersion along the main symmetry lines of the SBZ. Moreover, we discuss
the effect of spin-orbit coupling on the energy splittings and the spin-polarization 
of the main surface states and resonances. Whenever possible, we compare the results with
previously studied heavy metals surfaces.
We find empty resonances, located below a gap similar to the L-gap of the
(111) fcc surfaces, that have a downward dispersion and cross the Fermi level,
similarly to the recently studied Os(0001) surface. Their spin polarization at the
Fermi level is similar to that predicted by the Rashba model, but the usual level
crossing at $\bar{\Gamma}$ is not found with our slab thickness.
Moreover, for selected states, we follow the spin polarization along the high 
symmetry lines, discussing its behavior with respect to ${\bf k}_{\parallel}$, the 
wave-vector parallel to the surface.
\end{abstract}

\maketitle

\section{Introduction}
\label{sec1}
The presence of electronic states localized in the last few atomic layers of a solid, namely 
surface states, can give surfaces properties different from the bulk.\cite{zangwill} Moreover, since
surfaces lack inversion symmetry, even non-magnetic (i.e. time-reversal invariant) materials can
have surface states with a non-vanishing spin polarization. Hence, surface 
states might be practically useful for instance in spintronics applications, and it is worthwhile to 
characterize them. 

The energy dispersion of surface states with respect to $\bm{k_{\parallel}}$, the wave-vector parallel to 
the surface, and in some cases also their spin polarization, have been analyzed for many surfaces of different materials, 
by both theoretical and experimental techniques (Density Functional Theory, DFT,\cite{{au111_first},{au_111},{spin_au},{au111},{KKR},{dirac},{ir111},{os0001}} and photoelectron spectroscopy, PES,\cite{{PES},{L_gap_PES_first},{L_gap_PES},{H_Pt},{La_Pt111},{spin_STM},{Ag_Pt111},{PES_DFT_Ir111},{Graphene_Ir111},{AlBr3_Ir111}} angular- and 
spin-resolved). For instance, for the heavy metal surfaces, the L-gap surface states are well known.\cite{{au111_first},{au_111},{spin_au},{au111},{KKR},{dirac},{ir111},{os0001},{L_gap_PES_first},{L_gap_PES},{spin_STM},{Graphene_Ir111}} They show a characteristic split 
parabolic energy dispersion, that can be interpreted by the Rashba model \cite{rashba} as a relativistic effect due 
to spin-orbit coupling.

Recently, we found theoretically that Os(0001) surface \cite{os0001} should host Rashba split surface states around
$\bar{\Gamma}$, below a gap similar to the L-gap of the (111) fcc surfaces, but with an inverted dispersion
as in Ir(111).\cite{ir111} Moreover, energy splittings due to spin-orbit coupling are present in other surface states 
of Os(0001), such as the $S_2$, $S_3'$, and $S_4$ states analyzed in Ref. \onlinecite{os0001}.

Re(0001) is another interesting surface used \cite{{Pd_Re},{Oxygen_Re},{Pd_Re_2}} both as a support 
for other metallic layers and as a reactive catalytic surface. Very recently, it has been shown 
that artificially constructed Fe chains on top of Re(0001) surface exhibit a spin spiral state. \cite{Fe_on_Re_exp} 
Though Re(0001) has been widely studied, surprisingly little information is available about its 
electronic structure. Being similar to the other surfaces discussed above, Re(0001) could have states similar to the Rashba 
split surface states with an inverted dispersion as, e.g., in Ir(111) and Os(0001). Also the other surface states
could be similar, but both the energy dispersion and their spin polarization, are poorly known.

In this work, we study by ab-initio techniques the electronic structure of Re(0001). We characterize 
its main surface states and follow, for the most interesting ones, the average direction of the spin
polarization as a function of $\bm{k_{\parallel}}$. About $\bar{\Gamma}$, we find Rashba-like states with
negative curvature, which cross the Fermi energy, and we characterize their spin texture at the Fermi level. 
We find also several states already familiar from the study of the other surfaces. We analyze the main surface 
states that appear in the electronic band structure: in particular, at variance with Au(111), Pt(111), and Ir(111),
but as in Os(0001), we do not find the $S_8$ Dirac-like surface states, studied in Ref. \onlinecite{dirac}.

The paper is organized as follows: in Section \ref{sec2}, we describe the methods and the computational
parameters. In Section \ref{sec3}, we present the Re(0001) electronic band structure and analyze the main
surface states and resonances. In Section \ref{sec4}, we discuss the spin polarization of some selected states and finally,
in Section \ref{sec5}, we present our conclusions.

\section{Method}
\label{sec2}
First-principle calculations were performed by means of DFT \cite{{HK},{KS}} within the 
Local Density Approximation (LDA) scheme, as implemented in the Quantum ESPRESSO \cite{{QE},{QE_2}} and 
\texttt{thermo\_pw}.\cite{thermo_pw} packages The Perdew and Zunger's \cite{PZ} parameterization for the exchange and
correlation energy is used. Spin-orbit coupling effects are included by using the Fully Relativistic (FR) PAW  
method,\cite{FR_PAW} with 5$d$ and 6$s$ valence electrons and 5$s$ and 5$p$ semicore states (Pseudopotential
Re.rel-pz-spn-kjpaw$\_$psl.1.0.0.UPF from pslibrary.1.0.0 \cite{{pslibrary},{pslibrary_2}}).
Calculations on the bulk system, were performed with an hexagonal close-packed (hcp) structure at the theoretical LDA lattice constants: 
$a=5.175$ a.u., $c=8.338$ a.u. ($c/a = 1.611$), which are respectively 0.8\% and 1\% smaller than experiment 
($a_{exp} = 5.217$ a.u., $c_{exp} = 8.425$ a.u.). \cite{COD} 
The surface has been simulated by both a 24-layers and a 25-layers slab perpendicular to the [0001] direction 
in order to check the stability of the results with respect to the breaking of the inversion symmetry. 
The slab replicas have been separated by a vacuum space of $44$ a.u.. The slab crystal structure has been obtained 
from the bulk, with a further relaxation along the [0001] direction, which has the most relevant effects on the 
first three atomic layers: in particular, the distance between the first two layers decreases by $5.4 \%$ with respect 
to the idealized interlayer distance in the bulk, while the distance between the second and the third layer increases 
by $2.9 \%$. At a first stage, we performed a calculation with a starting non-zero magnetization, but the 
self-consistent ground state of the slab ended up to be non magnetic. The pseudo wavefunctions are expanded 
in a plane waves basis set with a kinetic energy cut-off of 60 Ry, while the charge density with a cut-off of 400 Ry.
BZ integrations were performed using a shifted uniform Monkhorst-Pack \cite{k_grid} $\bf{k}$-point mesh of $16 \times 
16 \times 1$ points for the slab and $16 \times 16 \times 10$ points for the bulk. 
The presence of a Fermi surface has been dealt with by the Methfessel-Paxton method \cite{MP} with a smearing parameter 
$\sigma=0.02$ Ry. With these parameters the total energy is converged within $10^{-3}$ Ry and the crystal parameters 
within $10^{-3}$ $\AA$.
\begin{figure*}[]
\centering
\includegraphics[width=0.5\textwidth,angle=0]{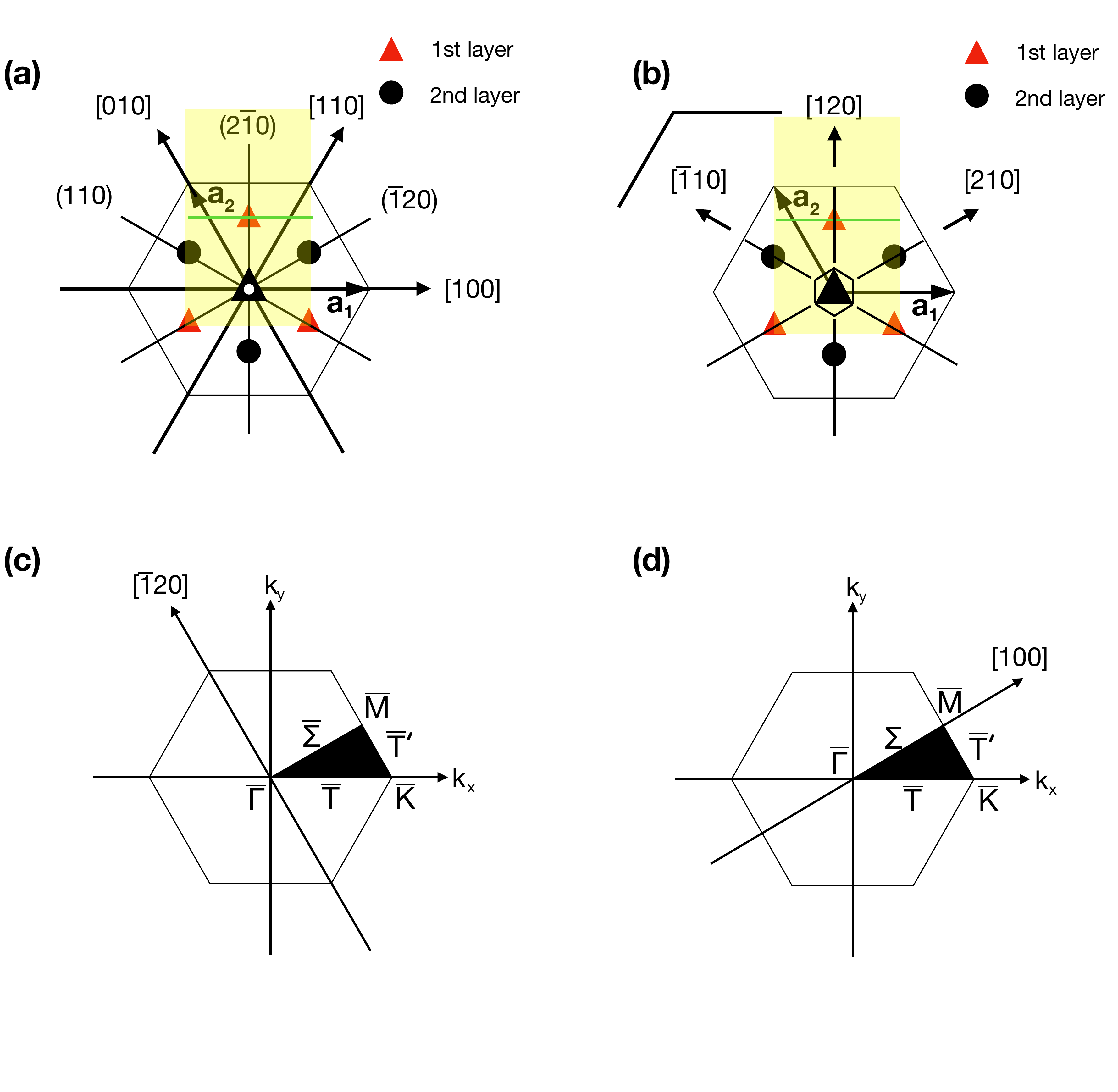}
\caption{(a)-(b) Positions of the atoms in the first two atomic layers of the Re(0001) 24-layers and 25-layers slab,
respectively. Arrows and solid lines indicate the $C_2$ rotation axes and the mirror planes, respectively.
(c)-(d) Surface Brillouin Zone of Re(0001) 24-layers and 25-layers slab, respectively. The Irreducible Brillouin Zone (IBZ) 
and the path used to plot the electronic band structure are shown. The $[\bar{1}20]$ and $[100]$ axes are the 
two-fold rotation axes of the small groups of $\bm{k_{\parallel}}$ along $\bar{T}'$ for the 24-layers and along 
$\bar{\Sigma}$ for the 25-layers slab, respectively.}
\label{f1}
\end{figure*}
\begin{figure*}[htbp]
\centering
\includegraphics[width=0.75\textwidth,angle=-90]{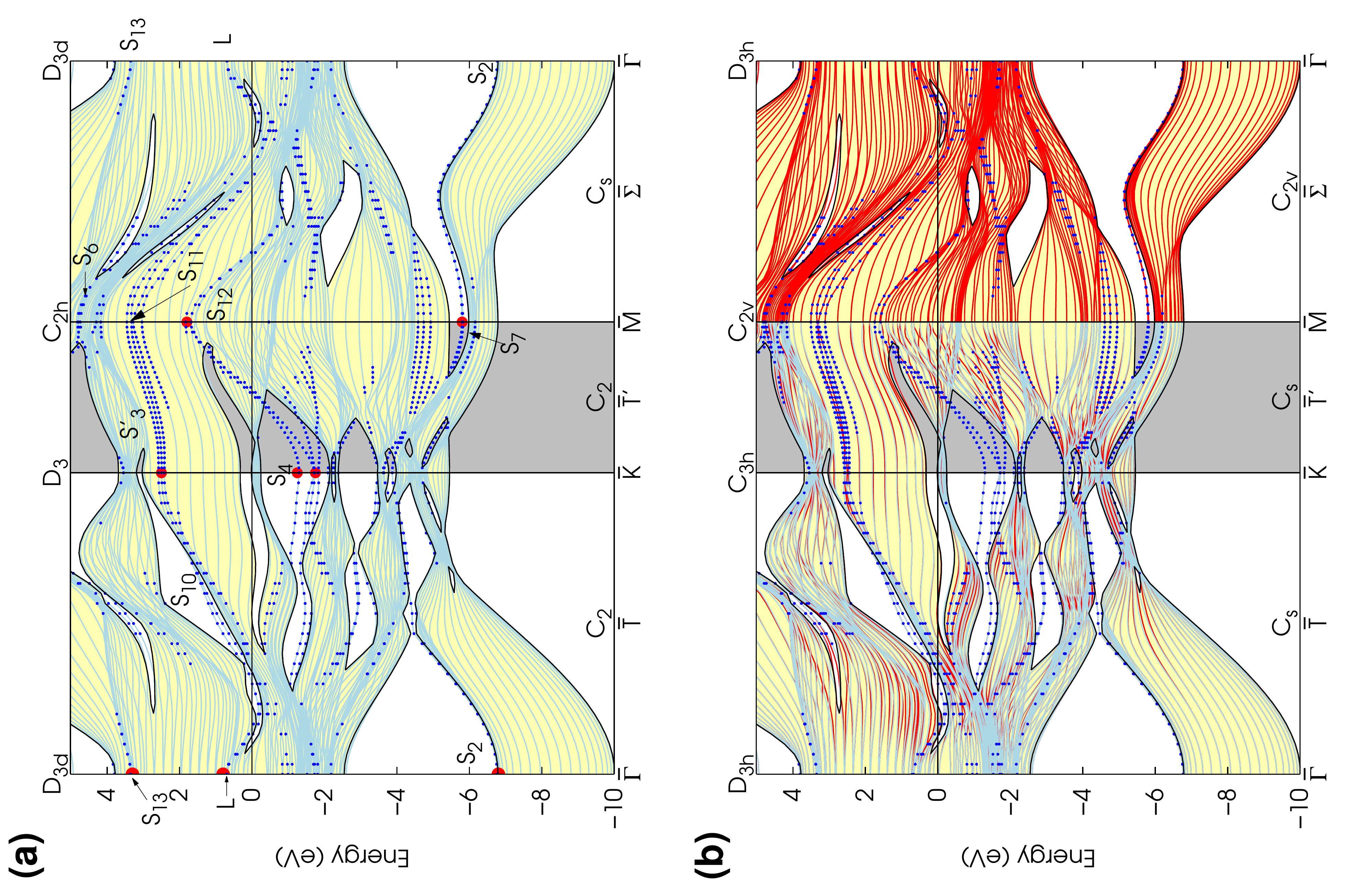}
\caption{(Color online) LDA FR-PAW surface band structure of Re(0001). (a) 24-layers slab band structure, (b) 25-layers slab band structure.
The yellow region is the Projected Band Structure (PBS), the light blue and red lines are the slab electronic states and the blue dots 
indicate surface states or resonances, defined as those having a charge density greater than 0.35 on the last 
two atomic layers of both surfaces. Energies are measured with respect to the Fermi energy, and the energy maximum 
in the figure is the computed work function (5.01 eV). Red dots in (a) indicate the states shown in Figs. \ref{f3} and \ref{f5}.}
\label{f2}
\end{figure*}
In Figs. \ref{f1}a-b we show the first two atomic layers of the 24-layers slab and the 25-layers slab, respectively. 
The 24-layers slab has a $D_{3d}$ point group. In particular, the $z$ axis, normal to the surface, is a $\bar{3}$ 
rotoinversion axis, while the axes $[100]$, $[110]$, and $[010]$ in Fig. \ref{f1}a are two-fold rotation axes. 
There are also three mirror planes, $(\bar{1}20)$, $(2\bar{1}0)$, and $(110)$ shown in Fig. \ref{f1}a. 
The 25-layers slab has instead, a $D_{3h}$ point group. The $z$ axis is a $\bar{6}$ axis, while the axes
$[210]$, $[120]$, and $[\bar{1}10]$, shown in Fig. \ref{f1}b, are two-fold rotation axes. Moreover, there are 
three mirror planes, whose traces coincide with the $C_2$ axes $[210]$, $[120]$, and $[\bar{1}10]$.
The electronic band structure was calculated along the path $\bar{\Gamma}-\bar{K}-\bar{M}-\bar{\Gamma}$ 
(that is along the $\bar{T}$, $\bar{T}'$, and $\bar{\Sigma}$ high-symmmetry lines) of the 
Surface Brillouin Zone (SBZ), shown in Figs. \ref{f1}c-d.
The small point group of $\bf{k}$ of the two slabs is indicated in the band structure in Figs. \ref{f2}a and \ref{f2}b, 
both for the high symmetry points ($\bar{\Gamma}$, $\bar{K}$, and $\bar{M}$) and for the high symmetry lines 
($\bar{T}$, $\bar{T}'$, and $\bar{\Sigma}$).
In particular, for the 24-layers slab, at $\bar{\Gamma}$, $\bar{K}$, and $\bar{M}$ the small group of $\bm{k}$ is 
$D_{3d}$, $D_{3}$, and $C_{2h}$, respectively. Along the high symmetry lines $\bar{T}$, $\bar{T}'$, and $\bar{\Sigma}$ 
it is $C_2$, $C_2$, and $C_s$, respectively. Along $\bar{T}$ the rotation axis coincides with the $x$-axis, while along
$\bar{T}'$ the rotation axis is the $[\bar{1}20]$ axis, shown in Fig. \ref{f1}c. Finally, along $\bar{\Sigma}$ the trace 
of the mirror plane of $C_s$ is $\bar{\Sigma}$.
On the other hand, for the 25-layers slab, at $\bar{\Gamma}$, $\bar{K}$, and $\bar{M}$ the small group of $\bm{k}$ is
$D_{3h}$, $C_{3h}$, and $C_{2v}$, respectively, while along the high symmetry lines $\bar{T}$, $\bar{T}'$, and $\bar{\Sigma}$
it is $C_s$, $C_s$, and $C_{2v}$, respectively. In particular, along $\bar{T}$ and $\bar{T}'$ the mirror plane is $\sigma_h$.
The two slabs have more symmetry elements than the Re(0001) surface, since they have symmetry operations that
exchange the two surfaces. Removing these elements, the surface point group is $C_{3\text{v}}$, while the small groups of 
$\bm{k}$ are $C_{3\text{v}}$, $C_3$, and $C_s$ for $\bar{\Gamma}$, $\bar{K}$, and $\bar{M}$ respectively and $C_1$, $C_1$, 
and $C_s$ along $\bar{T}$, $\bar{T}'$, and $\bar{\Sigma}$. Actually, they are the same for both slabs.

\section{Results}
\label{sec3}

\begin{figure*}
\centering
\includegraphics[width=0.5\textheight,angle=-90]{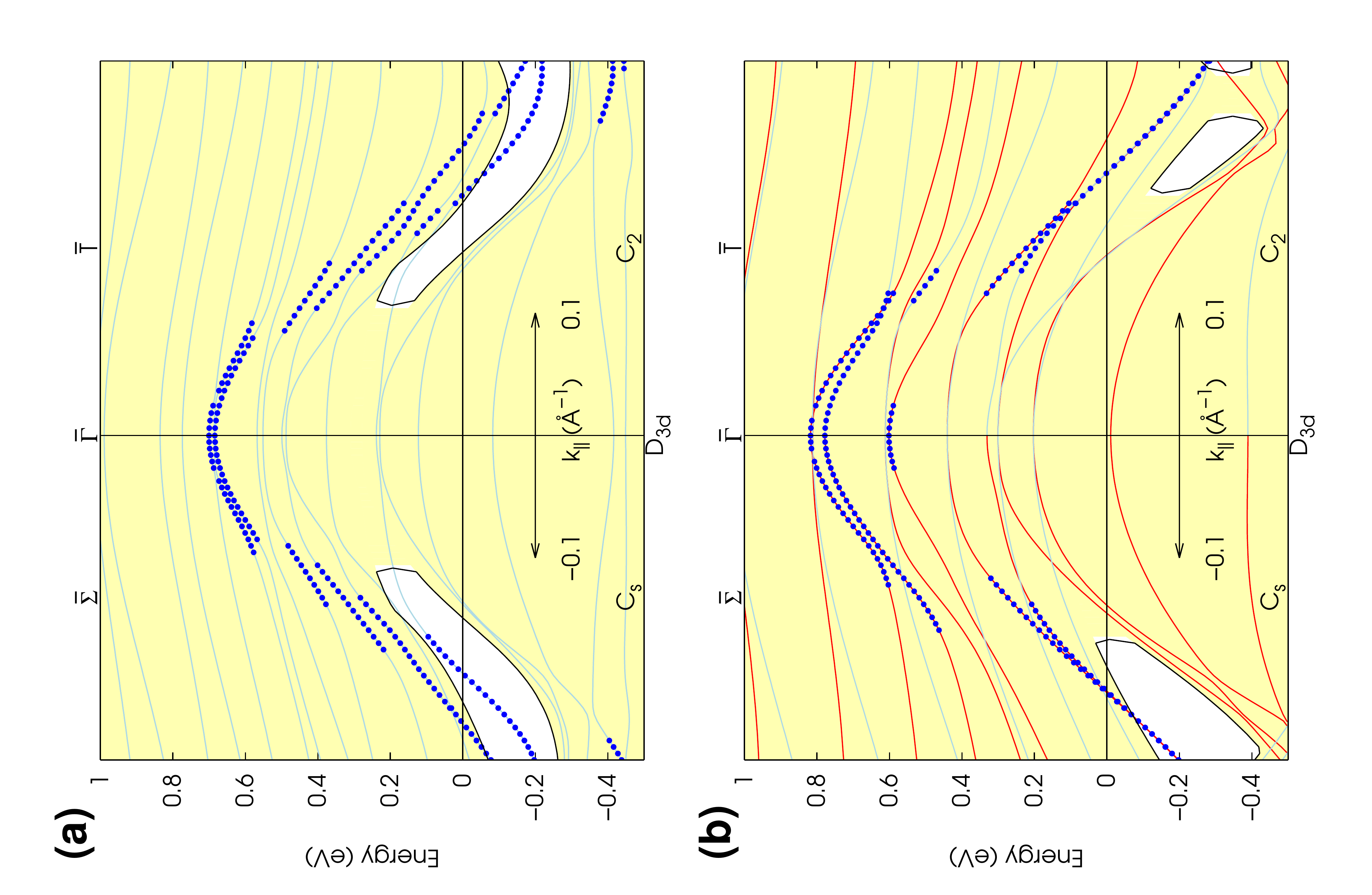}
\caption{Magnification of the electronic band structure around $\bar{\Gamma}$ in the FR (a) and SR (b) case. The $L$ states 
are shown with blue dots.} 
\label{f4}
\end{figure*}

In this section, we analyze the Re(0001) 24-layers slab FR band structure, shown in Fig. \ref{f2}a. We characterize the 
main surface states, indicated with red dots in Fig. \ref{f2}a, and compare them with Os(0001) and other previously studied surfaces (e.g. Au(111), Pt(111), and
Ir(111)). We use the same names as in Ref. \onlinecite{os0001}.
Moreover, at the end of the section we discuss the band structure of a 25-layers slab (Fig. \ref{f2} b).

\begin{table*}[]
\centering
%\scriptsize
%\setlength{\tabcolsep}{4pt} 
\resizebox{\textwidth}{!}{
\begin{tabular}{lcccccccccccc}
%\begin{tabular*}{\textwidth}{@{} l @{\hspace*{1.5mm}} c  @{\hspace*{1.5mm}} c  @{\hspace*{1.5mm}} c  @{\hspace*{1.5mm}} c  @{\hspace*{1.5mm}} c  @{\hspace*{1.5mm}} c  @{\hspace*{1.5mm}} c  @{\hspace*{1.5mm}} c  @{\hspace*{1.5mm}} c  @{\hspace*{1.5mm}} c  @{\hspace*{1.5mm}} c  @{\hspace*{1.5mm}} c  @{\hspace*{1.5mm}} c}
\hline
\hline
Surface   & ${\bf k}_{\parallel}$ & $\varepsilon$ (eV) & $\varepsilon$ (eV)& $\varepsilon$ (eV)&$\varepsilon$ (eV)&$\varepsilon$ (eV)& & Small & & &Symmetry & \\
State   & & Re(0001) & Os(0001)  & Ir(111) & Pt(111) & Au(111) & & group of $\bm{k}_{\parallel}$ & & & &  \\
\hline
L          & $\bar \Gamma$ & $0.68$ & $0.08$ & $-0.31$ & $0.1$ & $-0.5$ & $D_{3d}$ & $[D_{3h}]$ &  $(C_{3v})$ & $\Gamma_4^-, \Gamma_4^+$ & $[\Gamma_7, \Gamma_8]$ & $(\Gamma_4)$ \\
$S_{2}$    & $\bar \Gamma$ & $-6.77$ & $-7.87$ & $-8.0$ & $-7.4$ & $-7.6$ & & & & $\Gamma_4^+, \Gamma_4^-$ & $[\Gamma_7, \Gamma_8]$ & $(\Gamma_4)$ \\
$S_{13}$   & $\bar \Gamma$ & $3.35$ & $1.55$ & --- & --- & --- & & & & $\Gamma_4^+, \Gamma_4^-$ & $[\Gamma_7, \Gamma_8]$ & $(\Gamma_4)$ \\ 
$S'_{3a}$   & $\bar K$      & $2.58$ & $1.07$ & --- & --- & --- & $D_3$ & $[C_{3h}]$ & $(C_3)$  & $\Gamma_5\oplus\Gamma_6$ & $[\Gamma_{11}\oplus\Gamma_{12}]$ & $(2\Gamma_6)$ \\
$S'_{3b}$   & $\bar K$      & $2.49$ & $0.97$ & ---  & --- & --- & & & & $\Gamma_4$ & $[\Gamma_7\oplus\Gamma_9]$ & $(\Gamma_4\oplus\Gamma_5)$ \\
$S'_{3c}$   & $\bar K$      & $2.42$ & --- & ---  & --- & --- & & & & $\Gamma_4$ & $[\Gamma_7\oplus\Gamma_9]$ & $(\Gamma_4\oplus\Gamma_5)$ \\
$S_{4a}$   & $\bar K$      & $-1.30$ & $-2.35$ & $-2.7$ & $-2.8$ & $-3.7$ & & & & $\Gamma_4$ & $[\Gamma_8\oplus\Gamma_{10}]$ & $(\Gamma_4\oplus\Gamma_5)$ \\
$S_{4b}$   & $\bar K$      & $-1.72$ & $-2.74$ & $-3.1$ & $-3.1$ & $-4.0$ & & & & $\Gamma_4$ & $[\Gamma_7\oplus\Gamma_9]$ & $(\Gamma_4\oplus\Gamma_5)$ \\
$S_{4c}$   & $\bar K$      & $-1.73$ & $-2.72$ & $-3.1$ & $-3.1$ & $-4.0$ & & & & $\Gamma_5\oplus\Gamma_6$ & $[\Gamma_{11}\oplus\Gamma_{12}]$ & $(2\Gamma_6)$  \\
$S_{4d}$   & $\bar K$      & $-1.82$ & $-2.99$ & $-3.5$ & $-3.7$ & $-4.7$ & & & & $\Gamma_4$ & $[\Gamma_8\oplus\Gamma_{10}]$ & $(\Gamma_4\oplus\Gamma_5)$ \\
%$S_{6}$    & $\bar M$      & $$ & $2.93$ & $1.6$  & $0.6$ & --- & $C_{2h} (C_s)$ & $\Gamma_3^-\oplus\Gamma_4^- (\Gamma_3\oplus\Gamma_4)$ \\
%$S_{11}$    & $\bar M$      & $$ & $1.90$ & --- & --- & --- &  & $\Gamma_3^+\oplus\Gamma_4^+$, $\Gamma_3^-\oplus\Gamma_4^-\ (\Gamma_3\oplus\Gamma_4)$ \\
$S_{12}$    & $\bar M$      & $1.68$ & $0.86$ & --- & --- & --- & $C_{2h}$ & $[C_{2v}]$ & $(C_s)$ & $\Gamma_3^+\oplus\Gamma_4^+$ & $[\Gamma_5]$ & $(\Gamma_3\oplus\Gamma_4)$ \\
$S_{7}$    & $\bar M$      & $-5.80$ & $-7.00$ & $-6.7$ & $-6.3$ & $-6.6$ & & & & $\Gamma_3^+\oplus\Gamma_4^+, \Gamma_3^-\oplus\Gamma_4^-$ & $[\Gamma_5]$ & $(\Gamma_3\oplus\Gamma_4)$ \\
$S_{10}$   & $0.6\ \bar K$   & $1.42$ & $-0.24$ & $-0.8$ & $-1.2$ & --- & $C_2$ & $[C_s]$ & $(C_1)$ & $\Gamma_3\oplus\Gamma_4$ & $[\Gamma_3\oplus\Gamma_4]$ & $(\Gamma_2)$ \\
%$S_{12}$   & $0.6\ \bar M$   & $$ & $-1.90$ & $-2.6$ & $-2.6$ & --- & $C_s\ (C_s)$ & $\Gamma_3\oplus\Gamma_4$ \\
\hline
\hline
\end{tabular}}
\caption{Energy and symmetry properties of the surface states discussed
in the paper, for the Re(0001), Os(0001), Ir(111), Pt(111), and Au(111) surfaces. The reported
symmetry refers to the 24-layers slab. In square brackets, the symmetry for the 25-layers slab, 
in parentheses, the symmetry relevant for the surface.
}
\label{t1}
\end{table*}

\begin{figure*}
\centering
\includegraphics[width=17cm,angle=0]{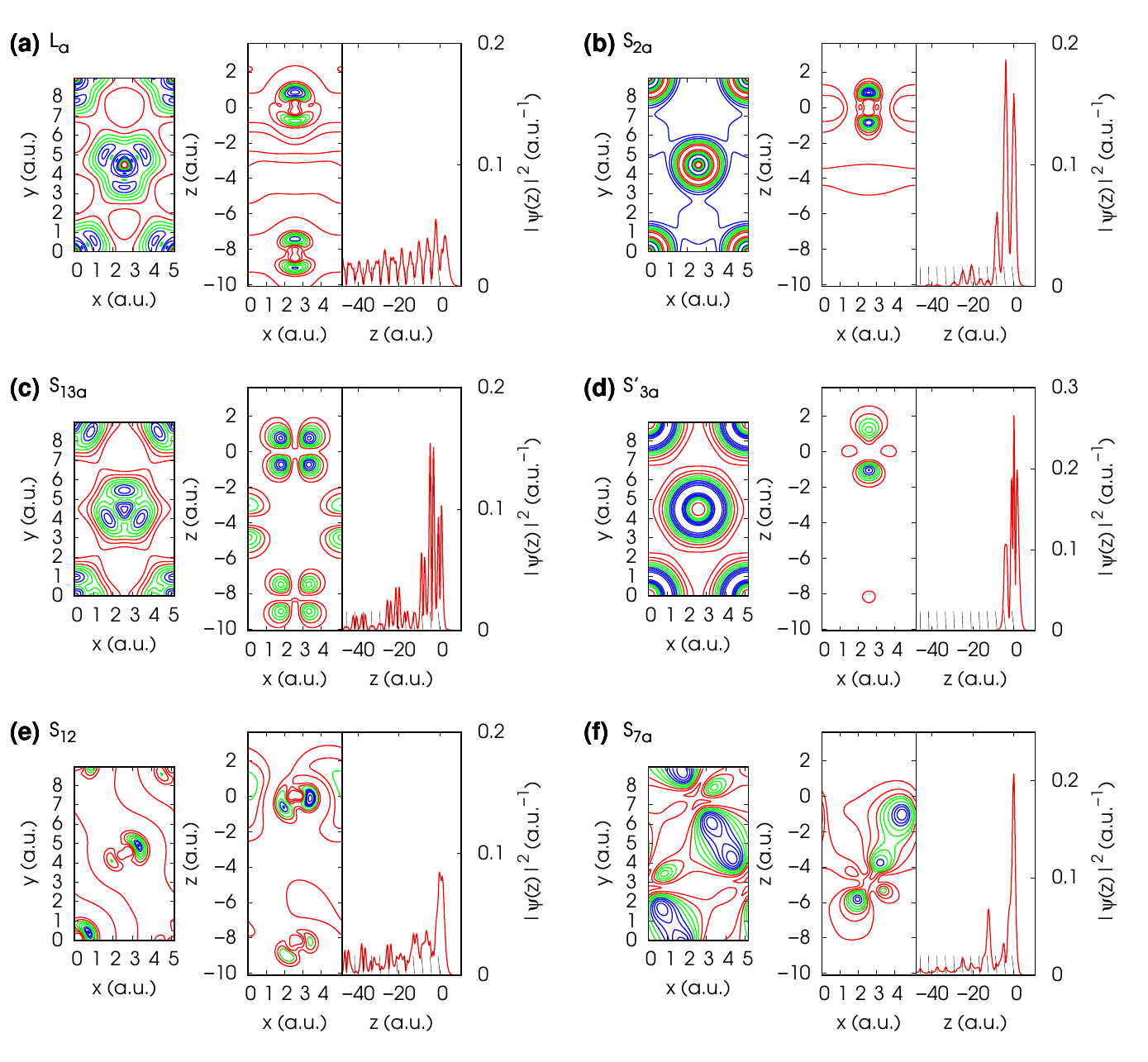}
\caption{Contour plots and planar average of the charge density corresponding to the selected FR surface states indicated with red dots in Fig. \ref{f2}.
The left subplot shows the charge density contour plot in the yellow region in Fig. \ref{f1}a, on the top atomic layer of the slab. The central subplot 
shows the contour plot in a plane perpendicular to the slab, whose trace is the green line
in Fig. \ref{f1}a. The contours are equally spaced and are indicated with different colors (red, green, and blue in increasing order of charge density).
The first three atomic layers are shown. The right subplot shows the planar average of the charge density 
in one half of the slab. The vacuum is on the right; the $z$ tics represent the positions of the atomic layers.}
\label{f3}
\end{figure*} 

We start our analysis from the $\bar{\Gamma}$ point, where we find two gaps in the PBS. Taking the Fermi energy as a
reference, the first is located 4 eV above it and the second approximately from $-7$ eV to $-3$ eV. The first gap, 
higher in energy, is similar to the L-gap of the fcc surfaces and is found in Os(0001) as well. It extends partly along
the $\bar{T}$ and $\bar{\Sigma}$ lines. The second gap, deeper in energy, extends up to half of the $\bar{T}$ line and
along the whole $\bar{\Sigma}$ line. 
Similarly to Os(0001) and Ir(111), but at variance with Au(111) and Pt(111), no surface states are found in the L-gap.
Below the L-gap, near the Fermi energy, we found two couples of states ($L$ in Fig. \ref{f2}a) that transform as the 
$\Gamma_4^-$ and $\Gamma_4^+$ representations of the $D_{3d}$ group. Their energy dispersion around $\bar{\Gamma}$ is parabolic with negative 
curvature, as for the Rashba split states in Os(0001) and Ir(111). In these surfaces we could fit their dispersion with 
the equation:
\begin{equation}
\centering
E_{\pm} = \frac{\hbar^2}{2 m^*} k_{\parallel}^2 \pm \gamma_{SO} k_{\parallel},
\label{eq1}
\end{equation}
where $k_{\parallel}$ is the modulus of the wave-vector parallel to the surface, $m^*$ is the effective electron mass and 
$\gamma_{SO}$ is the spin-orbit coupling parameter. However, here their dispersions do not cross at $\bar{\Gamma}$, as shown 
in Fig. \ref{f4} a, and even neglecting this splitting it is not possible to fit them with Eq. \eqref{eq1}. 
Nevertheless, at the Fermi energy the two states show a splitting along $\bm{k_{\parallel}}$, due to spin-orbit coupling: indeed,
a comparison with the Scalar Relativistic (SR) band structure (Fig. \ref{f4} b), shows that this splitting emerges only in 
the FR picture. Moreover, the spin texture of the $L$ states at the Fermi energy is well predicted by the Rashba model 
(see Section \ref{sec4} for more details), so they behave as Rashba states.
In Fig. \ref{f3}a we show the contour plots and the planar average of the sum of the charge densities at $\bar{\Gamma}$ 
for the $L_a$ states, the couple higher in energy. The contour plots suggest that it has mainly $s$ character hybridized
with some $d$ states. The planar average is maximum around the surface and shows a very slow decay towards the center of the
slab, indicating that the $L$ states are resonances. The gap at $\bar{\Gamma}$ could have several causes: 
among them, the evident hybridization with bulk states, possibly together with the finite size of the slab. Yet, a 
calculation with a 40-layers slab shows that the gap at $\bar{\Gamma}$ between $L_a$ and $L_b$ is the same as for the 
24-layers slab, thus finite-size effects do not play a relevant role in this case.

At lower energies at $\bar{\Gamma}$, there are two couples of states in a PBS gap, similar to the previously studied $S_2$ 
states of the other metal surfaces. At $\bar{\Gamma}$, they have symmetry $\Gamma_4^+$ and $\Gamma_4^-$. Their energy dispersion
has a positive curvature and can be fitted with \eqref{eq1}, with: 
$\gamma_{SO}= (0.200 \pm 0.005) \times 10^{-9}$ eV cm and $m^*/m = (0.661 \pm 0.003)$, with identical values, within the error bar,
along $\bar{T}$ and $\bar{\Sigma}$. In particular, $\gamma_{SO}$ is 30 \% lower than in Os(0001), while the effective mass is 
approximately 10 \% lower. The charge density contours and planar average of the $S_{2a}$ states, those higher in energy, 
are shown in Fig. \ref{f3}b. The states are surface states mainly localized on the first two atomic layers.

Finally, at $\bar{\Gamma}$ there are two couples of empty localized surface states called $S_{13}$, which have been characterized in
Os(0001) surface. At $\bar{\Gamma}$ they transform as the $\Gamma_4^+$ and $\Gamma_4^-$ representations of the $D_{3d}$ group. Similarly to
Os(0001), they are resonances and they have mainly $d$ character, with main contributions from the first two atomic layers (Fig. 
\ref{f3}c).

The states $L$, $S_2$, and $S_{13}$ extend partially also along the $\bar{T}$ line, where they all transform as the $\Gamma_3 \oplus \Gamma_4$
representation of the $C_2$ group. Along $\bar{T}$ we find some PBS gaps as well: the widest ones host the $S_{10}$, $S_4$, and 
the previously mentioned $S_2$ states. The $S_{10}$ states are two couples of degenerate states with symmetry $\Gamma_3 \oplus \Gamma_4$. They 
cross the Fermi level around $k_{\parallel} = 0.51 \, \AA^{-1}$. As in Pt(111), Ir(111), and Os(0001), they merge with the $S'_3$ states at $\bar{K}$. The 
$S_4$ states are located inside a PBS gap, they cross the $\bar{K}$ point and extend along the $\bar{T}'$ line as well. They
have symmetry $\Gamma_3 \oplus \Gamma_4$.

\begin{figure*}
\centering
\includegraphics[width=17cm,angle=0]{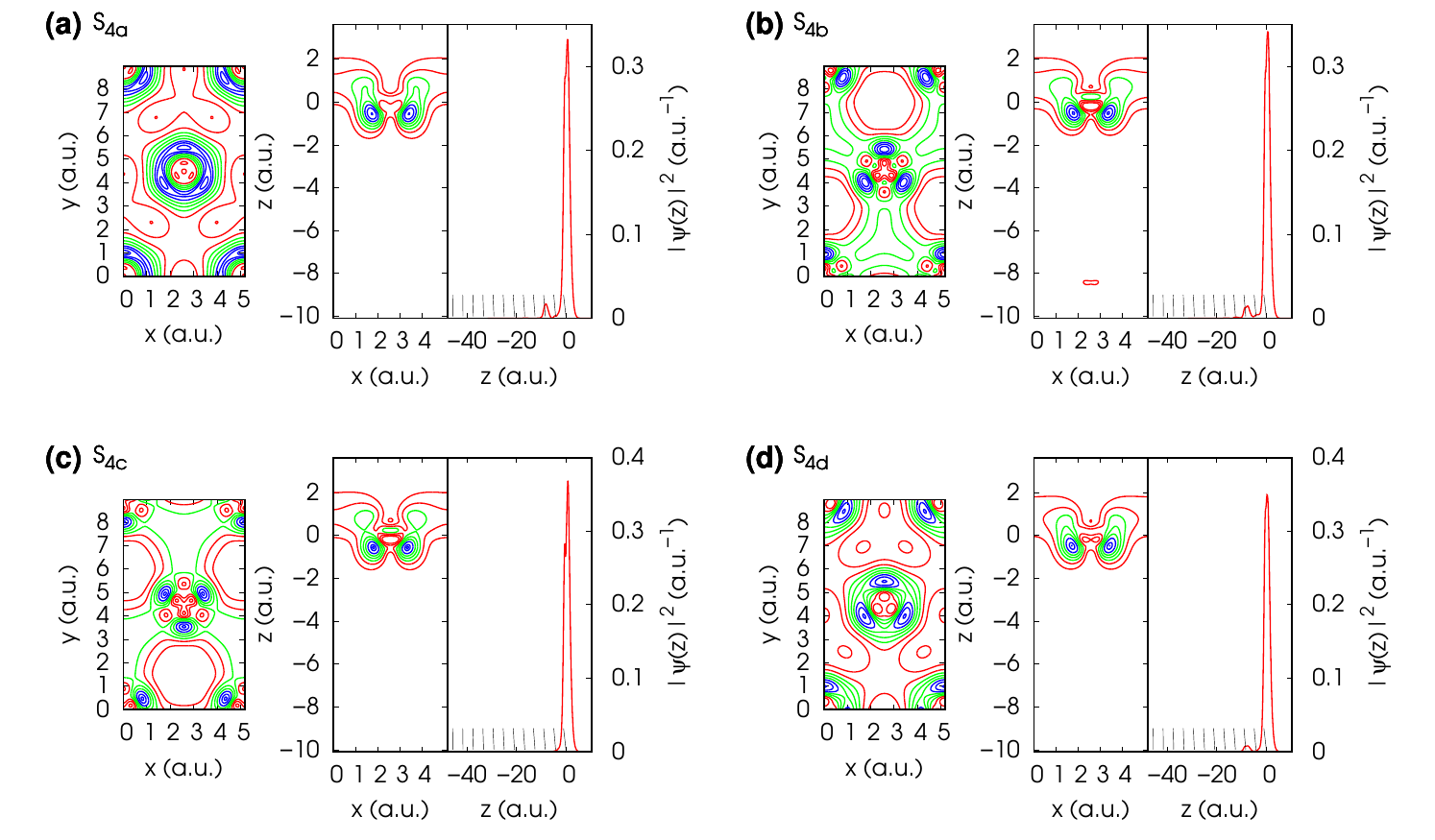}
\caption{Contour plots and planar average of the charge density of the $S_4$ surface states at $\bar{K}$. 
The organization of the subplots is the same as in Fig. \ref{f3}.}
\label{f5}
\end{figure*}

At $\bar{K}$ we find four main gaps in the PBS: the highest in energy is located above $3.5$ eV, the second one crosses the Fermi 
level and does not host any surface state, the third one contains the $S_4$ states, while the fourth one extends down to $-3.5$ eV. 
The main surface states at $\bar{K}$ are the $S'_3$ and $S_4$ states. $S'_3$ are made up of three couples of empty states, 
that are named $S'_{3a}$, $S'_{3b}$, and $S'_{3c}$ in decreasing order of energy. $S'_{3a}$ transforms as the $\Gamma_5 \oplus 
\Gamma_6$ representation of the $D_3$ group, while $S'_{3b}$ and $S'_{3c}$ have symmetry $\Gamma_4$. As in Os(0001), they are not 
in a PBS gap, they are localized in the first two atomic layers and project mainly on $d_{3z^2-r^2}$ states, as can be seen from
the charge density contour lines shown in Fig. \ref{f3}d.

The $S_4$ states are located in the PBS gap found at $-2 $ eV $< E < -0.3 $ eV. $S_{4a}$, $S_{4b}$, and $S_{4d}$ transform as the 
representation $\Gamma_4$, while $S_{4c}$ transforms as $\Gamma_5 \oplus \Gamma_6$. In Fig. \ref{f5} we show their charge density,
which is peaked on the top atomic layer, with a small contribution on the third atomic layer for $S_{4a}$, $S_{4b}$, and $S_{4d}$.
They have very similar features in Os(0001), though $S_{4b}$ and $S_{4c}$ exchange their character, 
as can be argued from the symmetry and the charge density plots. 

The PBS gaps and surface states described at $\bar{K}$ extend also along the $\bar{T}'$ line. Both $S'_3$ and $S_4$ states, along
$\bar{T}'$ have symmetry $\Gamma_3 \oplus \Gamma_4$ of the $C_2$ group. Moreover, $S_{4b}$ and $S_{4c}$ anticross near $\bar{K}$ 
($\bm{k} \approx 1.53 \, \AA^{-1}$), similarly to Os(0001). Along $\bar{T}'$, near $\bar{M}$, we find another PBS gap, located 
around $-6$ eV, that hosts the $S_7$ states. It extends up to $\bar{M}$ and along the whole $\bar{\Sigma}$ line, as well as the $S_7$ 
states, that connect to the $S_2$ states at $\bar{M}$.

At $\bar{M}$, besides the previously mentioned PBS gap and $S_7$ states, we find the $S_{12}$ states. They are a couple of degenerate
states with symmetry $\Gamma_3^+ \oplus \Gamma_4^+$ and project on many $d$ states (Fig. \ref{f3} e). The $S_7$ states, instead, 
are made up of two couples of states, that belong to the representations $\Gamma_3^+ \oplus \Gamma_4^+$ and $\Gamma_3^- \oplus \Gamma_4^-$ 
of the group $C_{2h}$, respectively. They have a strong contribution to the charge density (Fig. \ref{f3}f) coming from $d_{x^2-y^2}$ and
$d_{xy}$ orbitals localized in the first atomic layer. 

The band structure of the 25-layers slab (Fig. \ref{f2} b) is overall very similar to the one of the 24-layers slab and the surface states
are located at the same energies in both slabs. Nevertheless there are minor differences, due to the different symmetries of the two
slabs. In particular, since the 25-layers slab lacks inversion symmetry (its point group is $D_{3h}$), only the $\bm{k}$ - $-\bm{k}$ Kramers
degeneracy remains, and a spin splitting may appear, along some lines. This is the case of the lines $\bar{T}$ and $\bar{T}'$, in which states 
of different symmetry (in our case, even and odd with respect to the mirror plane $\sigma_h$) are split. The spin splitting is different for 
different states: it can be very small as, e.g., $\approx 10^{-6}$ eV for the $S_4$ states, or larger as $\approx 0.03$ eV for the $S'_3$ states, 
and it decreases increasing the slab thickness.\cite{41_layers} At variance with the states along $\bar{T}$ and $\bar{T}'$, the states along 
$\bar{\Sigma}$ are doubly degenerate because the $C_{2v}$ double group has only one two-dimensional irreducible representation, $\Gamma_5$.

\section{Spin polarization: results and discussion}
\label{sec4}
In this section we discuss the spin polarization of some of the surface states found above. 
The spin polarization can be obtained integrating the planar average of the magnetization density over half slab:
\begin{equation}
\centering
m_{\alpha}=\int_{0}^{L/2} m_{\alpha} (z) \, dz \, ,
\label{eq6}
\end{equation}
where the zero of $z$ is taken at the center of the slab and $L$ is its length along $z$, including vacuum. $m_{\alpha} (z)$ in Eq. \ref{eq6} is
the planar average of the magnetization density $m_{\bm{k}n}^{\alpha} (\bm{r})$ associated to the Bloch state 
$\braket{\bm{r} | \Psi_{\bm{k} n \sigma}}$ and is defined as:
\begin{equation}
\centering
m_{\alpha}(z) = \int_{A} m_{\bm{k}n}^{\alpha} (x,y,z) \, dx \, dy \, ,
\label{eq5}
\end{equation}
where $A$ is the yellow shaded region shown in Fig. \ref{f1}a, and 
\begin{equation} 
\centering
m_{\bm{k}n}^{\alpha} (\bm{r})= \mu_B \sum_n \sum_{\sigma_1,\sigma_2} \braket{\Psi_{\bm{k} n \sigma_1} | \bm{r}} \sigma^{\sigma_1 \sigma_2}_{\alpha} \braket{\bm{r} | \Psi_{\bm{k} n \sigma_2}},
\label{eq2}
\end{equation}
where $\mu_B$ is the Bohr magneton and $\sigma_{\alpha}$ are the Pauli matrices. The sum over $\sigma_1$ and $\sigma_2$ is over the spin,
while the sum over $n$ is over degenerate states (see Ref. \onlinecite{os0001} for more details).

\begin{figure*}
\centering
\includegraphics[height=\textwidth,angle=-90]{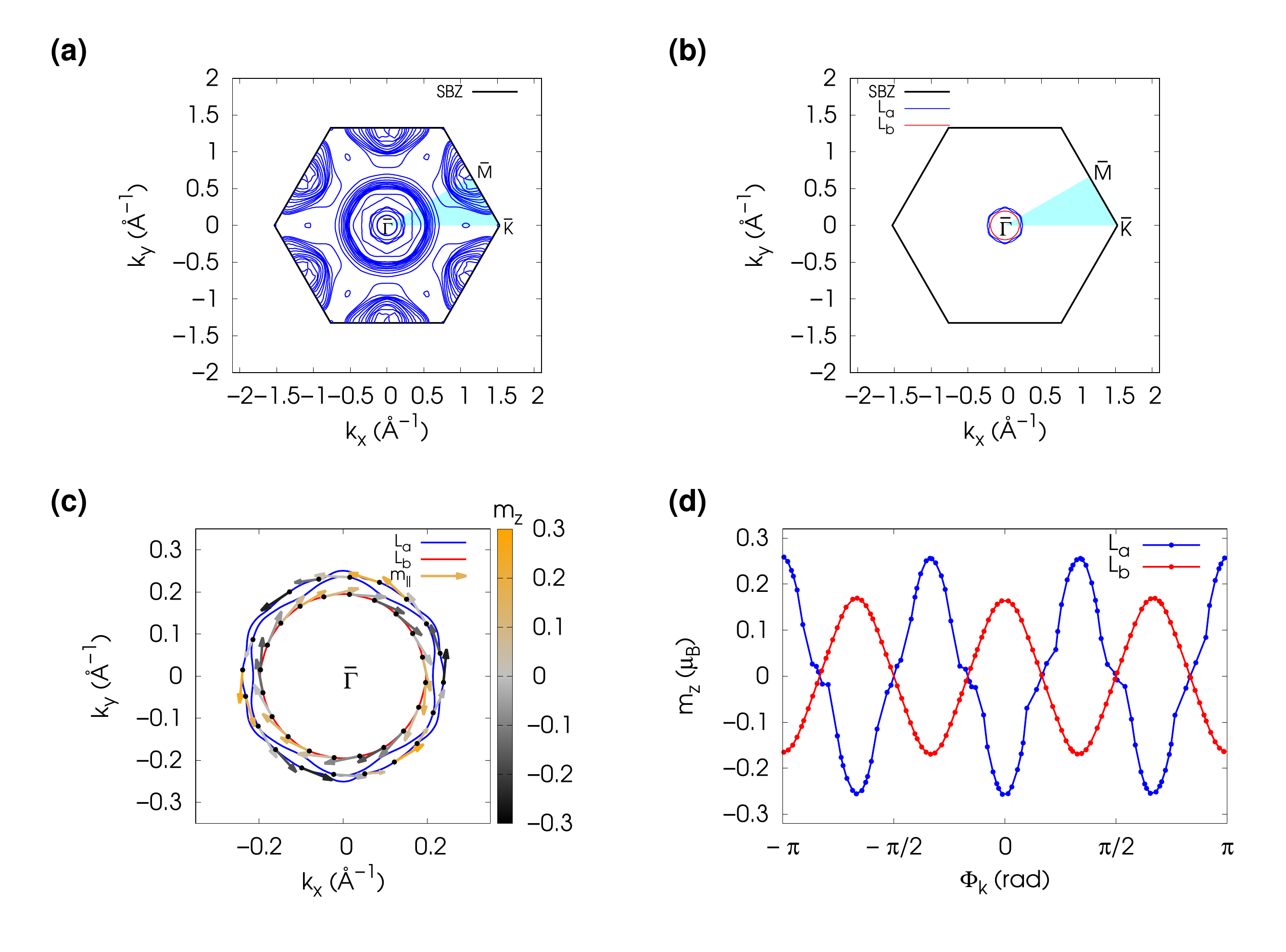}
\caption{(a) Fermi surface of the Re(0001) 24-layers slab. The light blue region is the Irreducible Brillouin Zone (IBZ). (b) The $L_{a,b}$ surface states 
contours shown in comparison with the SBZ. (c) Magnification of the $L_{a,b}$ surface states contours at the Fermi energy. The black 
dots indicate the surface states, the arrows indicate the spin polarization parallel to the surface, and they are colored depending on 
the magnitude of the $z$ component of the spin polarization. (d) $z$ component of the spin polarization for the states $L_{a,b}$ as a 
function of $\Phi_k = \text{tan}^{-1}(k_y/k_x)$.}
\label{f6}
\end{figure*}

We start our discussion from the $L$ states. In particular, we consider their contribution to the Fermi surface and their spin texture at 
the Fermi energy. The results are shown in Fig. \ref{f6}. The Fermi surface of the slab is shown in Fig. \ref{f6}a, while the contour 
levels of the $L$ states, shown in Fig. \ref{f6}b (compared with the SBZ) are magnified in Fig. \ref{f6}c: they have been obtained with a 
cubic interpolation of the energies of the states computed in a $14 \times 14$ square mesh of $\bm{k}$ points centered in $\bar{\Gamma}$. 
The spin polarization, computed via Eq. \eqref{eq6}, is represented by arrows whose length is proportional to the component of the 
spin parallel to the surface. The arrows are colored according to the magnitude of the spin polarization (Eq. \ref{eq6}) perpendicular 
to the surface, as indicated by the color map in the Figure. The $L_b$ states have a circular Fermi surface, whereas the shape of the $L_a$ 
states is more influenced by the underlying lattice. The component of the spin polarization parallel to the surface is perpendicular to the 
wavevector for both states, and it rotates clockwise and counter-clockwise for the two states, respectively. This is in agreement with the 
prediction of the Rashba model,\cite{rashba} so the $L$ states appear as Rashba split states at the Fermi level, although it has not been 
possible to fit their energy dispersion with Eq. \eqref{eq1}. In particular, given the dependence of the Rashba spin texture on the sign of 
both the effective mass and the spin-orbit coupling parameter,\cite{sign_rashba} our results are consistent with a Rashba model with 
$\gamma_{SO} > 0$. Due to the presence of the underlying atomic layers, the spin polarization shows a non vanishing component perpendicular 
to the surface. As shown in Fig. \ref{f6}d, this component oscillates around zero along the contour levels, with a period of $2 \pi /3$ as a 
consequence of the symmetry of the lattice, with opposite phase for $L_a$ and $L_b$. Similar effects can be 
simulated also in the Rashba model by introducing hexagonal warping effects.\cite{{warping},{warping_2}}

\begin{figure*}
\centering
\includegraphics[height=0.5\textwidth,angle=-90]{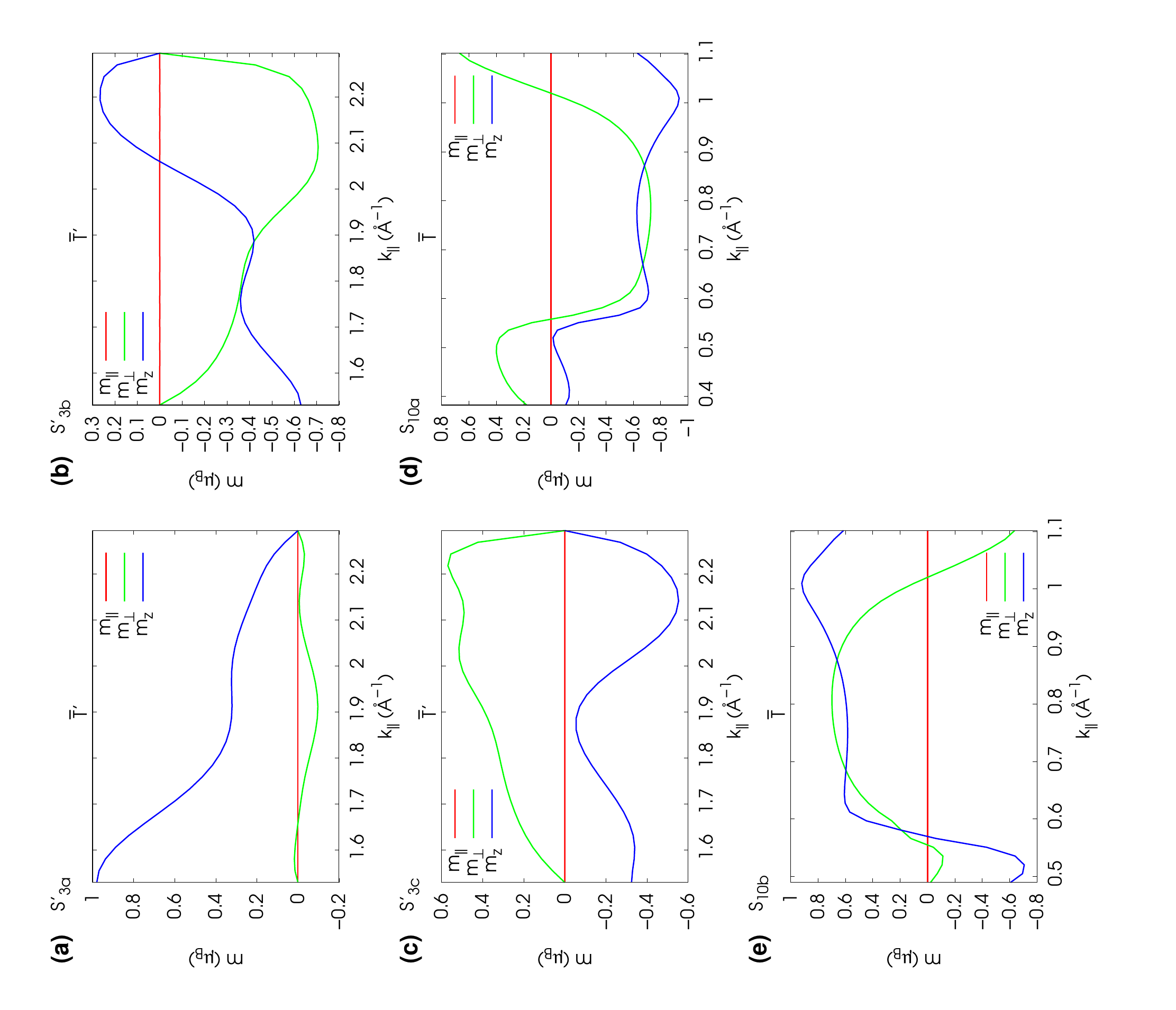}
\caption{Spin polarization components as a function of $\bm{k_{\parallel}}$ for the FR surface states $S'_{3a,b,c}$ and 
$S_{10a,b}$. $m_{\parallel}$ and $m_{\perp}$ are the spin polarization components parallel to the surface: they are parallel
 and perpendicular to the high symmetry line, respectively. $m_z$ is the component perpendicular to the surface.}
\label{f7}
\end{figure*}

\begin{figure*}
\centering
\includegraphics[height=0.5\textwidth,angle=-90]{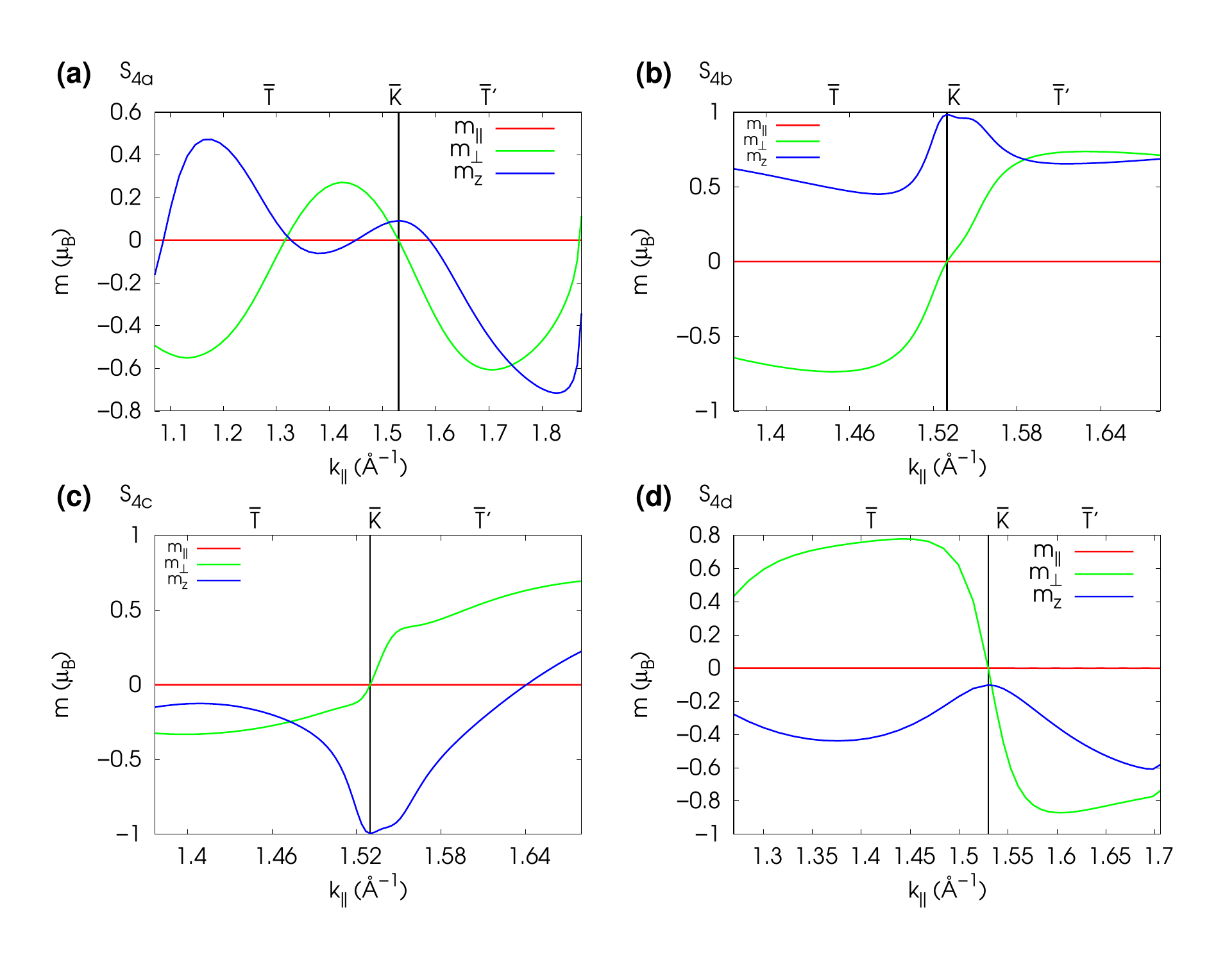}
\caption{Spin polarization components as a function of $\bm{k_{\parallel}}$ for the FR surface states $S_{4a,b,c,d}$. The convention on $m_{\parallel}$, $m_{\perp}$, and $m_z$ is the same as in Fig. \ref{f7}}
\label{f8}
\end{figure*}

Along the $\bar{T}$ and $\bar{T}'$ high symmetry lines the spin polarization can rotate in a plane perpendicular to the line, as 
explained in Refs. \onlinecite{os0001}, \onlinecite{symmetry_rashba}. In this work we consider the rotation of the spin polarization of the
states $S'_3$, $S_{10}$ (Fig. \ref{f7}), and $S_4$ (Fig. \ref{f8}).

The $S'_3$ states (Figs. \ref{f7}a-c) have been studied along the whole $\bar{T}'$ line: at $\bar{K}$ the states have only a non-zero 
$z$ (perpendicular to the surface) component of the spin polarization, due to symmetry constraints, while at $\bar{M}$ their spin 
polarization vanishes because $\bar{M}$ is a time-reversal invariant point. The spin polarization of $S'_{3a}$ is mainly perpendicular 
to the surface: the $z$ component decreases along the $\bar{T}'$ line, in a similar fashion as in Os(0001). The $S'_{3b}$ states show 
a more pronounced rotation: the $z$ component changes sign along the high symmetry line, and the component perpendicular to $\bar{T}'$ 
spans a wide range of values, at variance with Os(0001). Finally, the $S'_{3c}$ states have a rotating spin along $\bar{T}'$, which 
always points towards the center of the slab: its behaviour is similar to what shown in Os(0001).

The $S_{10}$ states show a smooth evolution of the spin polarization in the region $0.55 \, \AA^{-1} < k_{\parallel} < 1 \, \AA^{-1}$, as
shown in Figs. \ref{f7}d,e: in particular, $S_{10a}$ and $S_{10b}$ have opposite spin. Around $k_{\parallel} \approx 0.55 \AA^{-1}$ 
and $k_{\parallel} \approx 1 \AA^{-1}$ the spin polarization rotates more rapidly, because the two states anti-cross. Overall, their 
behaviour is similar to that shown by Os(0001).

Finally, the $S_4$ states (Fig. \ref{f8}) show a spin texture along $\bar{T}$ and $\bar{T}'$ very similar to Os(0001). In particular, 
the smoothest behaviour is shown by $S_{4d}$, for which the spin always points towards the slab. $S_{4b}$ and $S_{4c}$ have a rapidly 
varying spin, even in a very narrow range of $k_{\parallel}$ as shown in Fig. \ref{f8} b-c, due to their mixing and anticrossing around 
$\bar{K}$: a comparison with Os(0001) shows that their features are exchanged, as pointed out by their symmetry (see Table \ref{t1}).

Similar calculations have been performed for the 25-layers slab as well. The results are very similar to those discussed 
above, in particular for the $S_4$ and $S_{10}$ states, which have the same energy dispersion in the two systems. Instead, 
the spin polarization of the $S'_3$ states shows a somehow different behavior, characterized by more rapid variations,
which might be due to the mixing of the states caused by their non-negligible spin splitting. However, as pointed out by 
remark \onlinecite{41_layers}, the spin splitting decreases, though slowly, with increasing slab thickness, so we expect a better 
agreement using a thicker slab.
  
\section{Conclusions}
\label{sec5}
We discussed the electronic structure of Re(0001) surface. We analyzed its main surface states and resonances, focusing on the contours
and planar average of their charge density.
At $\bar{\Gamma}$ we found a gap similar to the L-gap of the (111) fcc surfaces. Like in the recently studied Os(0001) and at variance
with other well known metal surfaces (e.g. Au(111)), this gap does not contain any surface state. Two states that cross 
the Fermi level, with the same nature as the L-gap surface state of Au(111), have been found: their spin texture at the 
Fermi energy is similar to the one predicted by the Rashba model, though the energy dispersion crossing predicted at 
$\bar{\Gamma}$ has not been found. Rashba split states are actually of interest because they can have 
relevant applications in spintronics \cite{{spintronics_review},{rashba_applications}} and recently their engineering 
has been discussed, for instance, in ferroelectric oxides.\cite{FE_rashba} 

We found $S_2$, $S'_3$, $S_4$, $S_6$, $S_7$, $S_{10}$, $S_{11}$, $S_{12}$, and $S_{13}$ states as in other surfaces. In particular, $S_2$ are Rashba split
states whose dispersion has been fitted with parameters $\gamma_{SO}= (0.200 \pm 0.005) \times 10^{-9}$ eV cm and $m^*/m = (0.661 \pm 0.003)$.
The $S_8$ Dirac-like states instead have not been found.

Along $\bar{T}$ and $\bar{T}'$ the spin polarization can rotate in a plane perpendicular to the high symmetry line: for the $S'_3$, $S_4$, and $S_{10}$ 
states we followed this rotation as a function of $k_{\parallel}$. Some of them, as the $S'_{3}$ states, show a smooth behavior, while others 
(e.g. $S_4$ and $S_{10}$) have a more rapidly varying spin polarization, due to the anti-crossing and mixing of the states. 

Compared to the recently studied Os(0001) surface, Re(0001) shows similar surface states and resonances, although they are higher in 
energy with respect to the Fermi level because of the lower number of electrons per atom. The main differences are found in the $L$ states, which
are more hybridized with the bulk, as shown by the planar average of the charge density. Minor differences can be observed in the spin
textures of the $S'_3$ and $S_4$ states: in particular, the $S_{4b}$ and $S_{4c}$ states are exchanged with respect to those of Os(0001),
as pointed out also by their symmetries.

Our work has been developed using the DFT-LDA scheme. The Kohn-Sham eigenvalues are different, in principle, from the quasi-particle energies, 
hence it might be necessary to compute many-body corrections for a more detailed comparison with experimental data. However, since these kinds of 
calculations are more computationally demanding, they are usually performed only when LDA is not enough to explain the experimental results. On 
the other surfaces, the main features of the bands, such as the presence or absence of L-gap states, are well predicted by DFT-LDA, while the 
exact energy positions of the surface states might have small shifts. To the best of our knowledge, 
there are no experimental data to compare with our results. We hope that our theoretical calculations could motivate 
ARPES measurements on this surface and, in case of discrepancies, other theoretical calculations.

\section*{Acknowledgments}

Computational facilities have been provided by SISSA through its Linux Cluster and ITCS and by CINECA through the SISSA-CINECA 2018 Agreement.


\begin{thebibliography}{00}
\bibitem{41_layers} A band structure calculation of a 41-layers slab shows that the spin splittings decrease, but quite slowly: for instance, the splitting of the $S'_3$ states decreases of about 30 \%.
\end{thebibliography}
\end{document}